# MAF-Net: Multiple attention-guided fusion network for fundus vascular image segmentation

Yuanyuan Peng[1,2*], Pengpeng Luan[1], Zixu Zhang[1]

[1]School of Electrical and Automation Engineering, East China Jiaotong University, Nanchang 330000, China

[2]School of Computer Science, Northwestern Polytechnical University, Xi'An 710000, China

Correspondence

School of Electrical and Automation Engineering, East China Jiaotong University, Nanchang 330000, China

✉Email:3066@ecjtu.edu.cn

Abstract

Accurately segmenting blood vessels in retinal fundus images is crucial in the early screening, diagnosing, and evaluating some ocular diseases, yet it poses a nontrivial uncertainty for the segmentation task due to various factors such as significant light variations, uneven curvilinear structures, and non-uniform contrast. As a result, a multiple attention-guided fusion network (MAF-Net) is proposed to accurately detect blood vessels in retinal fundus images. Currently, traditional UNet-based models may lose partial information due to explicitly modeling long-distance dependencies, which may lead to unsatisfactory results. To enrich contextual information for the loss of scene information compensation, an attention fusion mechanism that combines the channel attention with spatial attention mechanisms constructed by Transformer is employed to extract various features of blood vessels from retinal fundus images. Subsequently, a unique spatial attention mechanism is applied in the skip connection to filter out redundant information and noise from low-level features, thus enabling better integration with high-level features. In addition, a DropOut layer is employed to randomly discard some neurons, which can prevent overfitting of the deep learning network and improve its generalization performance. Experimental results were verified in public datasets DRIVE, STARE and CHASEDB1 with $F_1$ scores of 0.818, 0.836 and 0.811, and Acc values of 0.968, 0.973 and 0.973, respectively. Both visual inspection and quantitative evaluation demonstrate that our method produces satisfactory results compared to some state-of-the-art methods.

## 1. Introduction

The blood vessels in retinal fundus images play an important role in the initial screening and diagnosis of ocular diseases, as such diseases typically cause changes in retinal vascular morphology [1-3]. Retinal fundus images can non-invasively capture microvessels and furnish rich information on vascular features such as diameter, shape, and curvature, which is vital in the prevention, diagnosis, and treatment of ocular diseases in clinical practice [4,5]. As shown in Figure 1, although retinal fundus images are widely used by clinicians to identify certain illnesses, the clinical practice of manual segmentation of these images is exceptionally difficult due to low contrast, noise, and irregular illumination [6-9]. The subjectivity involved in the segmentation process leads to inconsistent results across different doctors, which may impair the clinician's diagnosis [10].

In the past decade, computer aided diagnosis technology has enabled fully automated segmentation of fundus vascular images. Recent advances in machine learning techniques, specifically deep neural networks, have led to remarkable breakthroughs in information science and image analysis [11]. Especially, AlexNet [12] emerged as a deep convolutional neural network, paving the way for numerous follow-up networks, including VGGNet [13], ResNet [14,15], and ResNeXt [16]. Residual connections facilitated training and increased efficiency. Unfortunately, some fine vessels in retinal fundus vascular images have shallow contrast and irregular appearance. Additionally, some images have chaotic and blurred backgrounds, which are obstacles to accurate segmentation [17]. As shown in Figure 1, some weak and thin structures are difficult to detect in fundus images. In recent years, Transformer [18-22] has been gradually replacing some convolutional neural networks in computer vision. Unlike the latter, Transformer's self-attention mechanism constructs global information of fundus vascular images, reducing image information loss. However, Transformer requires a longer build time for the global attention mechanism and assistance with achieving necessary fast segmentation of fundus vascular images. Hence, Transformer and convolutional neural networks are integrated to develop a simple, practical and efficient algorithm for fundus vessel segmentation. The main contributions of our works are as follows:

(1) The UNet framework [23] was improved by applying a lightweight neural network with four smaller encoder and decoder blocks. In addition, Batch Normalization (BN) [24] and DropOut [25] were imported during convolution to prevent the improved neural network from overfitting.

(2) To avoid mutual interference, a self-attention fusion module is applied in the encoder stage, which aggregates spatial-channel attention units in parallel. This self-attention mechanism extracts global information from fundus images that compensates for the convolution's limitations.

(3) By introducing a spatial attention mechanism in the skip connection part, it filters out redundant information, assigns more weights to relevant information, and avoids interference from irrelevant knowledge.

(4) The merits of the deep learning model, DropOut layer, Batch Normalization, dual-attention mechanism and spatial attention mechanism are tightly integrated to generate an efficient image processing application framework for fundus vessel segmentation.

The rest of this study is organized as follows. The related works for segmentation of fundus vascular images is introduced in Section 2. The MAF-Net is described in Section 3. Section 4 exhibits the visual inspections and quantitative evaluations of experimental results. In section 5, a fascinating discussion is given to analyze the advantages and disadvantages of the MAF-Net. And in section 6, a conclusion is made for the MAF-Net.

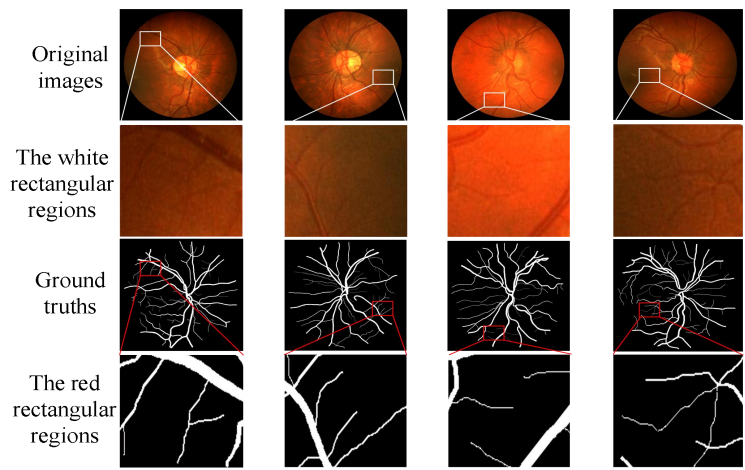

Figure1. Fundus vascular images.

## 2. Related works

Accurate segmentation of fundus images is important for the prevention and diagnosis of ocular diseases. To overcome the problem, many algorithms have been proposed for fundus vascular image segmentation. These segmentation methods can be roughly divided into three categories. The first category is to manually design the feature extraction layers to highlight the curvilinear structures, such as Hessian Matrix[26], second-order image derivatives[27], stick-based filters[28,29], and dynamic evolution models[30,31]. The second classification is the use of deep learning methods to segment fundus vascular images under different deep learning frameworks, such as FCN network [32], UNet network [23], CENet network [33], NAUNet network [34], ConvUNext network [35], CS2Net network [36], SA-Net network [37], and GDF-Net network [38]. The third strategy combines deep learning and manual design of feature layers to improve the segmentation accuracy of fundus vascular images, such as D-GaussianNet [39], the local intensity order transformation (LIOT) [40], and the combination of a ODoS filter and deep learning algorithms [41]. Although these methods have achieved good results, it is still a difficult task for weak fundus vascular images segmentation.

### 2.1 Traditional methods

The traditional method is to design feature extraction layers to extract feature information based on the unique shape and structure of fundus vascular images. Lesage et al.[26] proposed the Hessian matrix to precisely segment the curvilinear structures, but it may lead to breakage of the vessels. Using a different strategy, Li et al. [42] presented a hardware-oriented approach to enhance fundus vascular images, but the segmentation efficiency was not good for thin blood vessel images. However, the manual methods of designing feature layers could only extract a few features in images and could not accurately detect fundus vessels.

### 2.2 Deep learning based methods

In order to make up for the defect that traditional methods can not accurately extract fundus information, a large number of deep learning methods have been proposed to segment fundus vascular images. The Fully Convolutional Networks(FCN) [32] was the first deep neural network for images segmentation, but it may led to a loss of data and was quickly abandoned by researchers. Its improved version named UNet, by introducing a skip connection between the encoder and decoder to reduce information loss, but the information fusion in the skip connection was too rough. To solve the problem, Gu et al.[33] proposed the CENet, which used dilate convolution to enhance the extraction of images information and introduced DAC and ASPP to reduce the loss of information, but dilate convolution produced grid effect, which affected the segmentation of small blood vessels. Similarly, Han et al.[35] proposed the ConvUnext architecture with 7x7 convolution, which enhanced convolution field of perception and filtered out irrelevant information from the fundus images. Unfortunately, it was too computationally intensive and complex to meet the demand for fast segmentation. To save the time, Yang et al.[34] proposed a lightweight network NAUNet. It added a channel attention mechanism in the UNet model, which could able to segment fundus images quickly. Using the same strategy, Guo et al.[37] proposed SA-UNet architecture with a spatial attention mechanism to filter out irrelevant information, but it could not accurately segment weak blood vessels. Recently, inspired by the transformer model [18-22], Mou et al. [36] introduced CS2-Net to capture global information of fundus images. However, deep learning algorithms were mainly looking for powerful structures, ignoring the complexity of the algorithm.

## 2.3 The combination of traditional methods and deep learning algorithms

Recently, many deep learning algorithms have been combined with manually designed feature layers to achieve good results in fundus vascular image segmentation. Alvarado-Carrillo et al.[39] applied the combination of Gaussian filter and adaptive parameters to segment fundus images. Although good results were achieved, it consumed too much memory and time. Using a different strategy, Shi et al.[40] proposed a LIOT for detecting fundus vascular images, but it may result in information loss during the image transformation process. Recently, Peng et al.[41] improved the model [40] and combined ODoS filter with deep learning, achieving good results in fundus vascular image detection. However, these methods needed to adjust parameters for the most effective feature extraction. If the parameters were poor, the desired effect may not be achieved.

## 3. Method
## 3.1 Network architecture

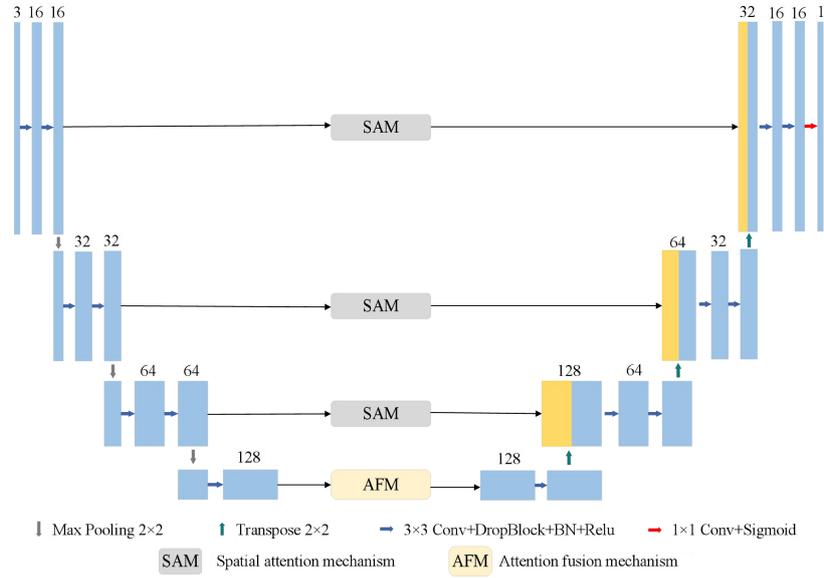

Figure 2. Network Architecture

The MAF-Net based on UNet framework is illustrated in Figure 2. The left encoder contains four convolution modules that are downsampled after each convolution, and an attention fusion mechanism (AFM) is introduced at the end of the encoder. The right-side decoder also includes four convolution modules, which are upsampled after each convolution. In the skip connection part, a spatial attention mechanism (SAM) is introduced to filter out redundant information. Finally, the segmentation map is obtained by outputting a 1x1 convolution and applying the Sigmoid function. In order to better understand the designed model, all details are presented in Table 1.

Table 1. The parameters details of the presented model

| Layers | Ouput size | Kernel size | Stride | Padding | Activation |
|---|---|---|---|---|---|
| Input Layer | 512×512×3 | - | - | - | - |
| Conv2D-1 | 512×512×16 | 3 | 1 | 1 | - |
| Dropout+BN-1 | 512×512×16 | - | - | - | ReLu |
| Conv2D-2 | 512×512×16 | 3 | 1 | 1 | - |
| Dropout+BN-2 | 512×512×16 | - | - | - | ReLu |

| Layer | Output Size | Kernel | Stride | Padding | Activation |
| --- | --- | --- | --- | --- | --- |
| SAM-1 | 512×512×16 | 7 | 1 | 3 | - |
| MaxPooling2D-1 | 256×256×16 | - | 2 | - | - |
| Conv2D-3 | 256×256×32 | 3 | 1 | 1 | - |
| Dropout+BN-3 | 256×256×32 | - | - | - | ReLu |
| Conv2D-4 | 256×256×32 | 3 | 1 | 1 | - |
| Dropout+BN-4 | 256×256×32 | - | - | - | ReLu |
| SAM-2 | 256×256×32 | 7 | 1 | 3 | - |
| MaxPooling2D-2 | 128×128×32 | - | 2 | - | - |
| Conv2D-5 | 128×128×64 | 3 | 1 | 1 | - |
| Dropout+BN-5 | 128×128×64 | - | - | - | ReLu |
| Conv2D-6 | 128×128×64 | 3 | 1 | 1 | - |
| Dropout+BN-6 | 128×128×64 | - | - | - | ReLu |
| SAM-3 | 128×128×64 | 7 | 1 | 3 | - |
| MaxPooling2D-3 | 64×64×64 | - | 2 | - | - |
| Conv2D-7 | 64×64×128 | 3 | 1 | 1 | - |
| Dropout+BN-7 | 64×64×128 | - | - | - | ReLu |
| Conv2D-8 | 64×64×128 | 3 | 1 | 1 | - |
| Dropout+BN-8 | 64×64×128 | - | - | - | ReLu |
| AFM | 64×64×128 | - | - | - | - |
| Conv2D-9 | 64×64×128 | 3 | 1 | 1 | - |
| Dropout+BN-9 | 64×64×128 | - | - | - | ReLu |
| Conv2DTranspose-1 | 128×128×64 | 3 | 1 | 1 | - |
| Concat(Conv2DTranspose-1,SAM-3) | 128×128×128 | 3 | 1 | 1 | - |
| Conv2D-10 | 128×128×64 | 3 | 1 | 1 | - |
| Dropout+BN-10 | 128×128×64 | - | - | - | ReLu |
| Conv2D-11 | 128×128×64 | 3 | 1 | 1 | - |
| Dropout+BN-11 | 128×128×64 | - | - | - | ReLu |
| Conv2DTranspose-2 | 256×256×32 | 3 | 1 | 1 | - |
| Concat(Conv2DTranspose-1,SAM-2) | 256×256×64 | 3 | 1 | 1 | - |
| Conv2D-11 | 256×256×32 | 3 | 1 | 1 | - |
| Dropout+BN-11 | 256×256×32 | - | - | - | ReLu |
| Conv2D-12 | 256×256×32 | 3 | 1 | 1 | - |
| Dropout+BN-12 | 256×256×32 | - | - | - | ReLu |
| Conv2DTranspose-3 | 512×512×16 | 3 | 1 | 1 | - |
| Concat(Conv2DTranspose-1,SAM-1) | 512×512×32 | 3 | 1 | 1 | - |
| Conv2D-13 | 512×512×16 | 3 | 1 | 1 | - |
| Dropout+BN-13 | 512×512×16 | - | - | - | ReLu |
| Conv2D-14 | 512×512×16 | 3 | 1 | 1 | - |
| Dropout+BN-14 | 512×512×16 | - | - | - | ReLu |
| OutLayer | 512×512×1 | 1 | 1 | - | Sigmoid |

**3.2 DropOut module**

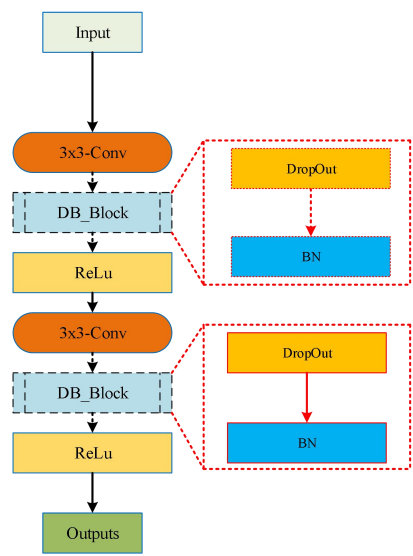

Figure 3. DropOut module

In spite of applying a data augmentation operation at the beginning of the presented framework, deep learning network shows overfitting while training retinal fundus images. To overcome the problem, the DropOut module is introduced to motivate the network to learn more robust and effective features [38]. Unlike the typical UNet convolution module, the DropOut and BN (Batch Normalization) are incorporated into the convolution process, as illustrated in Figure 3. This specific design module accelerates the convergence speed of the neural network while decreasing the model's overfitting problem.

When the model parameters of deep neural networks are huge while the training data is limited, the accuracy of the training data is high but the accuracy of the test data is meager. This overfitting hinders the neural network's convergence. Although data augmentation operation increases the number of images, it is still inadequate for neural networks. Specifically, the DropOut module is applied into the deep learning network to avoid overfitting phenomennon. In other words, the DropOut module is implemented randomly by deactivating some neurons during training. As a result, the reliance of neural networks on particular features is reduced, motivating them to learn more robust and generic features. In this way, the neural network overfitting can be prevented [25].

The incorporation of the BN layer has two principal advantages. Firstly, it can rectify the internal variable offset. During the training stage of the deep learning network, the distinctiveness of each layer's distribution exacerbates the complexity of learning. The integration of BN produces a data conversion in every layer that guarantees a mean of zero and variance of one. This operation facilitates the convergence of the model. Secondly, it can alleviate the problem of gradient disappearance and gradient explosion. As the deep neural network intensifies, gradient-related complications like gradient disappearance or explosion commonly occur and create an obstacle in training neural network models. The adoption of the BN layer mitigates this problem to a considerable extent [24].

**3.3 Attention fusion mechanism**

The use of dual-attention mechanism to enrich contextual information to compensate for the loss of scene information during downsampling was first introduced by Fu et al. [43] in a scene segmentation task. Inspired by these studies [37, 43], a self-attention fusion module is applied in this deep learning network.

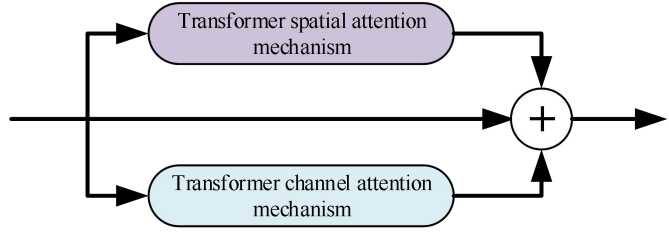

Figure 4. Attention fusion mechanism (AFM)

The local features acquired from the neural network may bring about classification errors, due to their localization properties that inhibit effective modeling of fundus images at a global level [44]. Drawing on the Transformer model, a dual-attention mechanism is introduced to capture fine-grained information in fundus images, as illustrated in Figure 4. It then combines them in parallel to minimize mutual disruption and fully extract detailed information of blood vessels in fundus images. The spatial attention mechanism influences long-range correlations and enables the model to obtain global information from the fundus images. In contrast, the channel attention mechanism primarily weighs the information, assigning greater importance to channels containing valuable data, while reducing the importance of channels with less pertinent information.

The global feature extraction issue can be addressed through the use of spatial attention mechanisms, which enhance the capacity of the model to learn the underlying global features. To strengthen the capturing of fundus vessels across the horizontal and vertical axis, the conventional convolution operation are replaced by a combination of 1x3 and 3x1 asymmetric convolution. As displayed in Figure 4, the mechanism receives input features F ∈ $R^{C \times H \times W}$, and linearly maps it to produce three matrices: Qy, Kx, and V=$R^{C \times H \times W}$. Qy and Kx correspond to vertical and horizontal directional features extracted from the fundus image, whereas C represents the number of feature channels in the input image, and H and W denote width and height, respectively. Upon remapping Qy, Kx, and V to RC×N, where N = H×W, Qy will be dotted with the transpose of Kx to generate a spatial attention feature map via Softmax, calculated as follows:

$$S(x,y) = \frac{\exp(Q_y^T \cdot K_x)}{\sum_{x'-1}^{N} \exp(Q_y^T \cdot K_{x'})} \tag{1}$$

where S(x,y) represents the influence of the pixel point at the yth position on the pixel point at the xth place, the spatial attention feature map can sufficiently learn the vascular structures at different spatial parts, with greater similarity contributing to higher feature map values. F' ∈ $R^{C \times N}$ denotes the spatial attention features generated pointwise by multiplying the feature matrix V with the spatial attention feature map. These spatial attention features F' are summed with the input vector F to extract contextual information from the fundus vascular image, ultimately improving the accuracy of image segmentation.

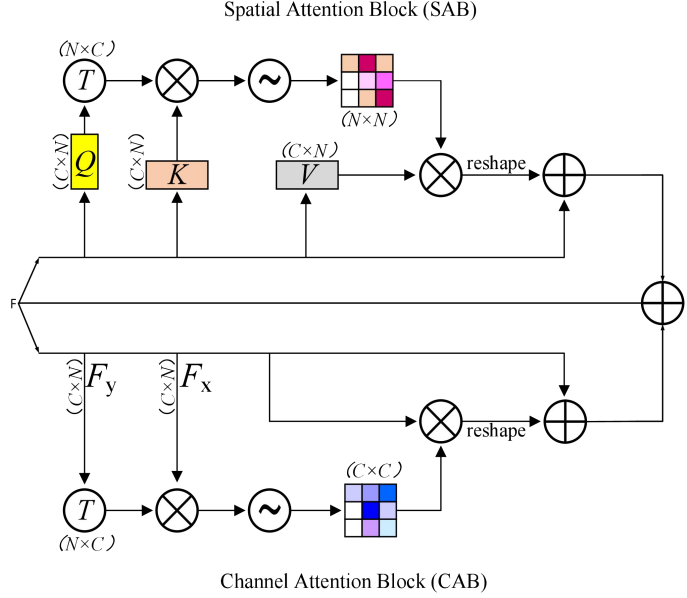

Figure 5. Self-attention-based spatial attention and channel attention [36]

The attention mechanism mimics human neural signal transmission. When humans view an image, they focus on a specific region based on their needs, and the information in other areas is automatically ignored. Therefore, the spatial attention mechanism is integrated with the channel attention mechanism to extract detailed information from fundus vessels in parallel. Classifying each channel of the fundus image into different classes may help establish interrelatedness among internal semantic features, thus drawing out relevant information from the channels to improve the semantic expression ability. The channel attention mechanism, as displayed in Figure 5, is computed using the Softmax function, which dots the input Fx ∈ $R^{C \times H \times W}$ with its transpose matrix Fy ∈ $R^{H \times W \times C}$ to generate the channel attention feature map. The detailed formula is demonstrated as follows:

$$C(x,y) = \frac{\exp(F_x \cdot F_y)}{\sum_{x=1}^{C} \exp(F_x \cdot F_y)} \qquad (2)$$

Here, C(x,y) denotes the attention of the xth channel relative to the yth channel. By computing the attentional feature map of the two channels, channels with high similarity are strengthened while channels with low similarity are suppressed. Then, the Softmax activation function is used to differentiate between background and vascular structures in the retinal fundus image. The channel attention feature map is point multiplied by the input features to get the channel attention features. Summing the channel attention features and the input features together enhances the contrast between different channels and improves the overall performance of the model.

### 3.4 Spatial attention mechanism

The UNet model compensates for downsampling by combining image information at the encoder side with image information. However, this approach mixes low-level features on the encoder side with high-level features on the decoder side, leading to minimal effectiveness. In fact, it introduces redundant and irrelevant information to high-level features, resulting in poor segmentation, specifically for fundus vessels [45, 46]. Inspired by these papers [35-38, 47], the spatial attention mechanism is introduced to skip connection to filter out redundant information and noise from low-level features, thus enabling better integration with high-level features. The principal strategy is illustrated in Figure 6. Through maxpooling and average pooling, matrix vectors $F_{maxpool} \in R^{H \times W \times C}$ and $F_{avgpool} \in R^{H \times W \times C}$ are generated. Subsequently, the spatial attention feature map is generated after convolution operation and Sigmoid. The filtered low-level features are obtained by the dot product of the attentional feature map and the input features. This mechanism improves the accuracy of segmenting fundus vascular images. The calculation process is as follows:

$$Fs = F \cdot \sigma(f^{7 \times 7}([MaxPool(F); AvgPool(F)])) \\ = F \cdot \sigma(f^{7 \times 7}[F_{maxpool}; F_{avgpool}]) \quad (3)$$

where $f^{7 \times 7}(\cdot)$ stands for convolution operation with convolution kernel 7, $\sigma(\cdot)$ denotes for Sigmoid function.

The spatial attention mechanism discussed in this section is distinct from the Transformer-based spatial attention mechanism described in Section 3.3, which primarily aims to achieve global modeling by extracting global information from fundus images to supplement the local information obtained through convolution [37]. Unlike the Transformer-based spatial attention, the spatial attention mechanism in this section does not rely on the self-attention mechanism that uses convolutional neural networks to eliminate irrelevant information in low-level features and enhance the fusion of low-level and high-level features.

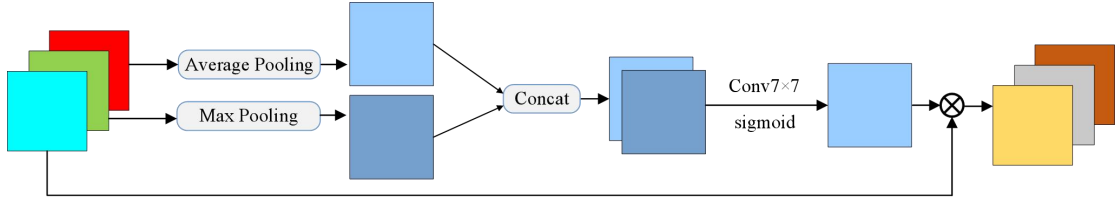

Figure 6. Spatial attention mechanism (SAM)

### 3.5 Loss function

The segmentation task for 2D fundus images can be regarded as a binary classification task at the pixel level, with retinal vessels and background as possible classes. In this type of task, a pixel-level loss function, such as binary cross entropy (BCE), may be used to assess neural network algorithms during training. BCE is capable of evaluating the similarity between true values and image labels. Mathematically, BCE is defined as follows:

$$\ell_{BCE} = -\frac{1}{N} \sum_{i=1}^{N} g_i \cdot \log(p_i) + (1 - g_i) \cdot \log(1 - p_i) \quad (4)$$

where $g_i \in \{0,1\}$ denotes the label value of the retinal fundus image, $p_i \in \{0,1\}$ means the predicted value of the retinal fundus image, and $N$ represents the pixel value.

## 4. Experiments
### 4.1 Datasets

DRIVE, STARE, and CHASEDB1 [36, 37] are retinal fundus image vascular detection datasets that are publicly available. DERIVE contains 40 images, which are split equally into training and testing images, has a resolution of 584 × 565 pixels. STARE includes 20 images for training and 10 images for testing, with a resolution of 605 × 700 pixels. CHASEDB1 consists of 28 images: 20 for training and 8 for testing. The resolution of the images is 999 × 960 pixels.

**4.2 Implementation details**

There are fewer publicly available datasets of fundus vascular images, which makes fewer fundus images available for training and tends to cause overfitting of the neural network. To avoid overfitting of the neural network, we introduce data augmentation strategy. Random rotation, adding Gaussian noise, and color jittering are used on retinal fundus images to increase the number of pictures. In addition, all the algorithms have the same epoch and learning rate. Where the epoch is 100 and the learning rate is 0.001. The details are shown in Table 2.

Table 2. The specific information of DRIVE, STARE, and CHASEDB1

| Datasets | DRIVE | STARE | CHASEDB1 |
|---|---|---|---|
| Implement / GPU | Pytorch / GeForce GTX3060 | | |
| Learning rate | 0.001 | | |
| Augmentation methods | (1) Random rotation ; (2) adding Gaussian noise; (3)color jittering; | | |
| Train / Test | 20 / 20 | 10 / 10 | 20 / 8 |
| Resolution(pixels) | 584×565 | 605×700 | 999×960 |
| Resize(pixels) | 592×592 | 512×512 | 1008×1008 |

**4.3 Evaluation Metrics**

Retinal fundus vessel segmentation involves a pixel-level binary classification task aimed at determining whether each pixel point in an image belongs to the positive or negative class. Here, positive pixels represent fundus vessels, while negative pixels represent other parts of the image. To evaluate the performance of the neural network's output in comparison to the true labels, true positive (TP), false positive (FP), false negative (FN), and true negative (TN) are computed in the confusion matrix. TP represents the number of pixels that correctly detect fundus vessels as fundus vessels, FP is the number of pixels that wrongly detect background as fundus vessels, TN is the number of pixels that correctly detect background as background, and FN is the number of pixels that wrongly detect the vessel class as background. These values are then used to calculate the evaluation metrics, including accuracy (Acc), sensitivity (SE), specificity (SP), and F1-score for vessel segmentation in retinal fundus images. Mathematically

$$Acc = \frac{TP+TN}{TP+FP+TN+FN} \tag{5}$$

$$SE = \frac{TP}{TP+FN} \tag{6}$$

$$SP = \frac{TN}{TN+FP} \tag{7}$$

$$F_1 = \frac{2 \times TP}{2 \times TP + FP + FN} \tag{8}$$

## 4.4 Visual inspection

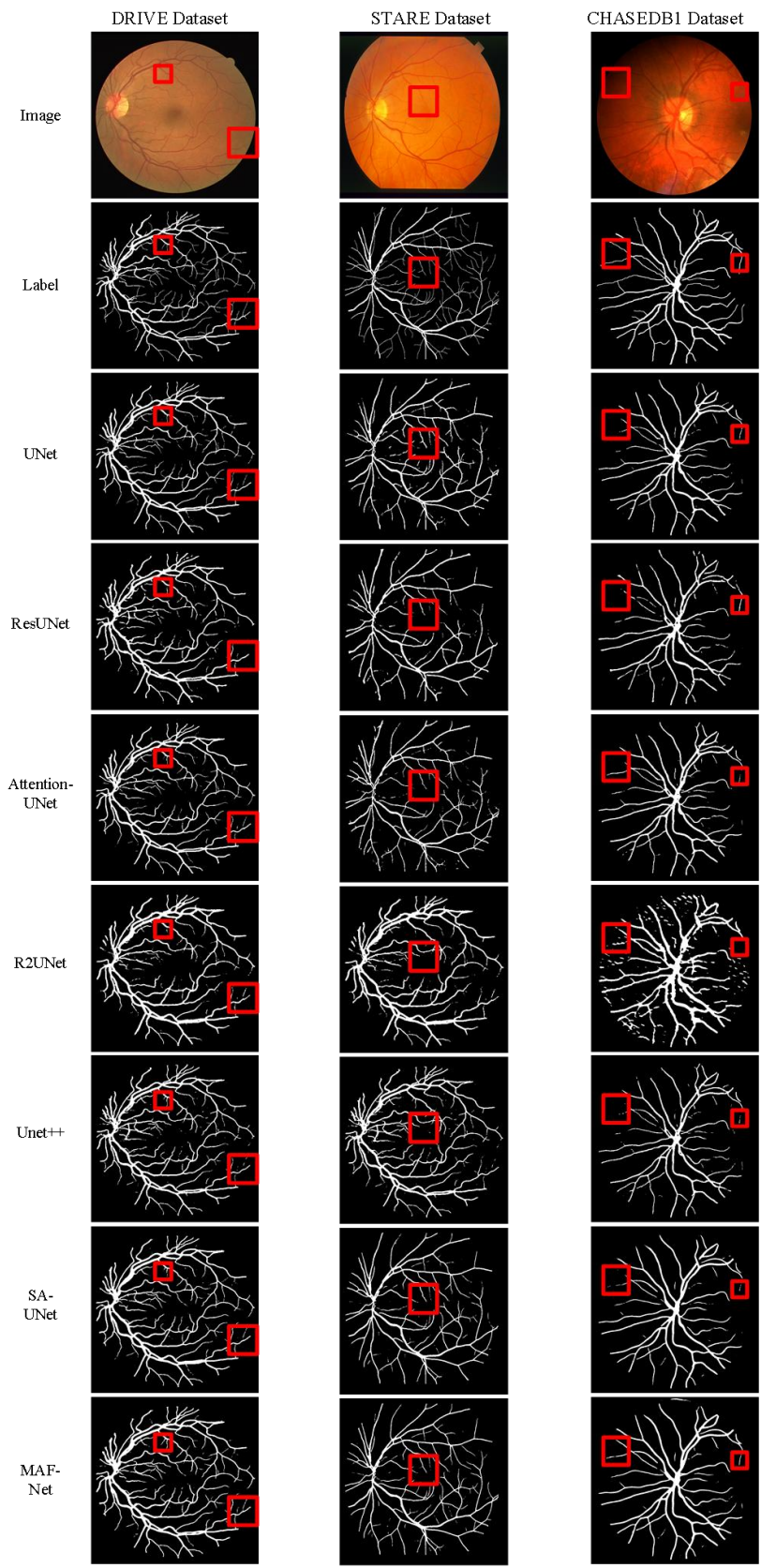

Figure 7. Experimental results with different methods were validated in DRIVE, STARE, and CHASEDB1 datasets.

In order to prove the effectiveness of the proposed algorithm, experimental results were conducted on three public datasets. As shown in Figure 7, the original image and the corresponding ground truth are given on the first and second lines. The segmentation results of the UNet [23], ResUNet [48], Attention-Unet [49], R2UNet [50], UNet++[51], SA-UNet [37] and MAF-Net are shown in the third, fourth, fifth, sixth, seventh, eighth and nineth lines, respectively. As can be seen from the red marks, the disadvantage of the comparison methods is that it cannot completely detect the weak vascular structures. The main reason is that the comparison methods cannot complete the interaction between the local information and the global information in the fundus images, leading to the loss of some details.

**4.5 Quantitative evaluation**

Table 3. Quantitative evaluation with different methods.

| Dataset | Methods | F1 | ACC | SE | SP |
|---|---|---|---|---|---|
| CHASEDB1 | UNet | 0.756 | 0.961 | 0.630 | 0.996 |
| | ResUNet | 0.723 | 0.956 | 0.593 | 0.995 |
| | Attention_UNet | 0.765 | 0.962 | 0.649 | 0.995 |
| | R2UNet | 0.698 | 0.939 | 0.740 | 0.960 |
| | UNet++ | 0.733 | 0.958 | 0.598 | 0.997 |
| | SA_UNet | 0.804 | 0.973 | 0.806 | 0.985 |
| | MAF-Net | **0.811** | **0.973** | **0.847** | 0.982 |
| Stare | UNet | 0.720 | 0.951 | 0.583 | 0.996 |
| | ResUNet | 0.759 | 0.956 | 0.642 | 0.994 |
| | Attention_UNet | 0.738 | 0.954 | 0.601 | 0.997 |
| | R2UNet | 0.732 | 0.950 | 0.629 | 0.989 |
| | UNet++ | 0.783 | 0.960 | 0.671 | 0.995 |
| | SA_UNet | 0.815 | 0.971 | 0.852 | 0.982 |
| | MAF-Net | **0.836** | **0.973** | **0.885** | 0.981 |
| DERIVE | UNet | 0.753 | 0.948 | 0.619 | 0.996 |
| | ResUNet | 0.736 | 0.945 | 0.608 | 0.994 |
| | Attention_UNet | 0.755 | 0.949 | 0.623 | 0.996 |
| | R2UNet | 0.761 | 0.948 | 0.653 | 0.991 |
| | UNet++ | 0.745 | 0.947 | 0.612 | 0.995 |
| | SA_UNet | 0.814 | 0.967 | 0.835 | 0.976 |
| | MAF-Net | **0.818** | **0.968** | **0.850** | 0.978 |

Table 4. Algorithm params and flops

| Algorithm | Swin-Unet | UNet Transforme | SA-UNet | Ours |
|---|---|---|---|---|
| Params(M) | 17.91 | 10.63 | 0.48 | 0.51 |
| Flops(G) | 60.92 | 414.24 | 16.24 | 16.5 |

As shown in Table 3 and Table 4, the proposed algorithm in this study has exhibited better performance than existing algorithms [23, 37, 48-51]. The MAF-Net demonstrated good results in handling thin and weak contrast fundus vascular images with uneven contrast. It can efficiently and accurately segment fundus vascular images while reducing the error rate, thus aiding clinicians in diagnosing some ocular diseases.

**4.6 Ablation study**
**4.6.1 The influence of each module on the presented framework**

Table 5. The influence of each module on the presented framework

|  | DB | SAM | AFM | F1 | ACC | SE | SP |
|---|---|---|---|---|---|---|---|
| **STARE** | × | √ | √ | 0.799 | 0.971 | 0.774 | **0.987** |
|  | √ | × | √ | 0.823 | 0.971 | **0.897** | 0.977 |
|  | √ | √ | × | 0.829 | 0.973 | 0.881 | 0.981 |
|  | √ | √ | √ | **0.836** | **0.973** | 0.885 | 0.981 |

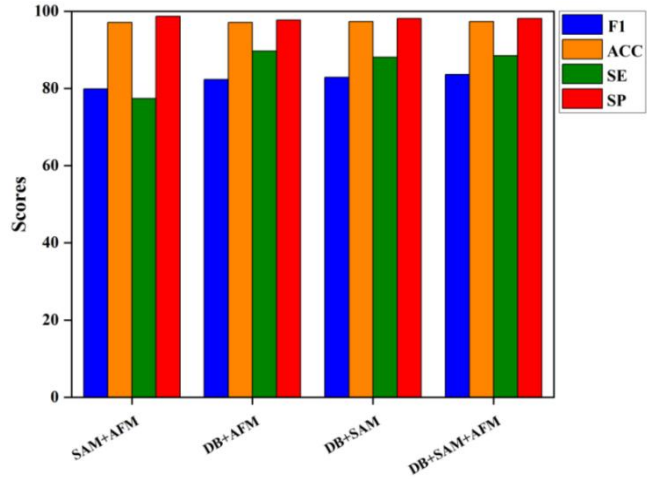

Figure 8. The influence of each module on the presented framework

To demonstrate the effectiveness of our proposed algorithm, ablation studies were conducted on the STARE dataset to verify the validation of different modules in the presented MAF-Net. We compare the DropOut layer with the BN layer (DB), the spatial attention mechanism (SAM), and the attention fusion mechanism (AFM). As shown in Table 5 and Figure 8, we verified the validity of each module to the algorithm in the STARE dataset. It can be seen that removing one module reduces F1 and Acc values of the entire model. This proves that the module improves the accuracy of fundus vascular image segmentation while increasing a small amount of computation.

**4.6.2 The effect of parameters in the DropOut layer on the presented framework**

Table 6. The influence of parameters in the DropOut layer on the presented framework

|  | DropOut | F1 | ACC | SE | SP |
|---|---|---|---|---|---|
| **STARE** | p=0.5 | 0.683 | 0.939 | 0.884 | 0.944 |
|  | p=0.7 | 0.725 | 0.949 | **0.904** | 0.953 |
|  | p=0.9 | **0.836** | **0.973** | 0.885 | **0.981** |

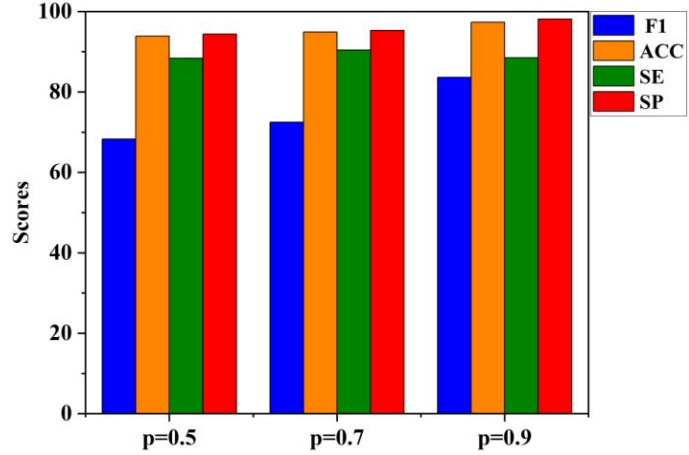

Figure 9. The influence of parameters in the DropOut layer on the presented framework

The DropOut layer can alleviate overfitting in the MAF-Net, but the parameters in DropOut also have a certain impact on neural networks. Therefore, we selected different parameters for validation on the STARE dataset, and the results are shown in Table 6 and Figure 9. Specifically, we set the parameters in DropOut to 0.5, 0.7, and 0.9 respectively to obtain experimental results. When the DropOut parameter is 0.9, F1 and Acc can reach their maximum values.

4.6.3 Comparison of state-of-the-art methods in the DRIVE dataset

Table 7. Compared with some state-of-the-art methods in the DRIVE dataset.

| Years | Method | ACC | SE | SP |
|---|---|---|---|---|
|  | MAF-Net | **0.968** | **0.850** | 0.978 |
| 2023 | Yang et al.[52] | 0.958 | 0.797 | 0.981 |
| 2023 | Liu et al.[53] | 0.956 | 0.799 | 0.979 |
| 2023 | Yang et al.[34] | 0.968 | 0.795 | 0.968 |
| 2022 | Dong et al.[54] | 0.959 | 0.795 | 0.921 |
| 2022 | Mubashar et al.[45] | 0.966 | 0.781 | 0.987 |
| 2021 | Guo et al.[37] | 0.967 | 0.835 | 0.976 |
| 2021 | Mou et al.[36] | 0.963 | 0.822 | 0.989 |
| 2020 | Saroj et al.[55] | 0.954 | 0.731 | 0.976 |
| 2020 | Palanivel et al.[56] | 0.948 | 0.736 | 0.979 |

To illustrate the validation of the proposed model, we compared with some state-of-the-art methods [23, 34, 36, 37, 45, 48-56]. As shown in Table 7, it can be seen that the MAF-Net has the highest ACC and SE values. This indicates that the MAF-Net for segmentation of fundus images is closer to real annotations. The main reason is that the MAF-Net can detect weak blood vessels in fundus images, resulting in the maximum SE value and ACC value.

5. Discussion

This study presents a unique method based on multiple attention mechanisms and deep learning for segmenting fundus vascular images, which performs well even in the presence of thin, weak, and inhomogeneous vessels. The presented MAF-Net framework has many specific characteristics and advantages. (1) The proposed MAF-Net framework improved the UNet model by importing the DropOut module and BN (Batch Normalization) module in the convolutional neural network. These modifications improve the network's performance while maintaining the segmentation accuracy. (2) To overcome the limitations of convolution and achieve global modeling of fundus vascular images, a dual-attention mechanism based on Transformer is introduced at the end of the encoder. This facilitates interaction with the local information derived by the convolutional neural network. Thereby strengthened parallel extraction of the information in the fundus vessel images, and ultimately improving the segmentation accuracy of tiny vessels. (3) In the skip connection part, a spatial attention mechanism is applied to eliminates noise and irrelevant information from the encoder image, ensuring that the decoder images are effectively combined to improve the accuracy of fundus vascular image segmentation. Accurate segmentation of blood vessels in fundus images can effectively assist doctors in diagnosing various eye diseases.

The MAF-Net was validated on three publicly available datasets: DRIVE, STARE, and CHASEDB1. Compared with some state-of-the-art methods [23, 34, 36, 37, 45, 48-56], both visual inspection and quantitative evaluation exhibit that the proposed algorithm has an excellent performance in thin, weak, and inhomogeneous vessel segmentation. The main reasons are as follows:(1) The dual attention mechanism can extract the information of fundus blood vessels from space and channel in the presented deep learning framework, respectively. (2) The spatial attention mechanism filters out noise from the encoder side so that the irrelevant information cannot carried into the decoder side. (3) Reducing the number of channels can accelerate the algorithm's running speed and improve its running speed, but the approach cannot reduce the segmentation accuracy of fundus vascular images.

However, due to the uneven contrast and significant changes in lighting in fundus vascular images, the proposed method may lead to breakage and incompleteness of the segmented small blood vessels. Additionally, the proposed method focus on developing deep architectures and ignore capturing the shape features of blood vessels in fundus images. Although the presented deep learning framework has many drawbacks, it can effectively detect weak vascular structures.

## 6. Conclusion

This paper aims to accurately segment blood vessels in retinal fundus images, particularly addressing the challenges of segmenting weak, thin, and inhomogeneous vessels. In which, DropOut and Batch Normalization are introduced to prevent the UNet model from overfitting. Additionally, since fundus vessel images contain detailed information which cannot be fully extracted by convolutional neural networks, the spatial attention mechanism and channel attention mechanism based on Transformer are presented to parallelly extract global knowledge from fundus vessel images. The loss of information in the downsampling stage is alleviated through feature fusion of the encoder and decoder. Unfortunately, the approach may result in noise and irrelevant information in the low-level features of the encoder to the decoder. The spatial attention mechanism is imported to alleviate the problem. The proposed method is validated on publicly available datasets, i.e., DRIVE, STARE, and CHASEDB1. Experimental results demonstrate that the presented MAF-Net framework is more successful when compared with some state-of-the-art methods [23, 34, 36, 37, 45, 48-56]. Although our algorithm can segment some weak, thin, and inhomogeneous fundus vessels, some vessels may still be break. Future research should consider how to address the issue of vessel breakage.

**CRediT authorship contribution statement**

**Yuanyuan Peng:** Conceptualization, Data curation, Methodology, Software, Validation, Visualization, Writing original draft. **Pengpeng Luan and Zixu Zhang:** Conceptualization, Funding acquisition, Writing – review& editing.

**Declaration of competing interest**

The authors declare that they have no known competing financial interests or personal relationships that could have appeared to influence the work reported in this paper.

**Data availability**

Data underlying the results presented in this paper are available in Ref. [36, 37].

**Acknowledgements**

This research was supported by the Jiangxi Provincial Natural Science Foundation (nos. 20212BAB202007, 20202BAB212004, 20224BAB202024).